# Gotta Assess 'Em All: A Risk Analysis of Criminal Offenses Facilitated through PokémonGO

ASHLY FULLER, University College London, United Kingdom

MARTIN LO, University College London, United Kingdom

ANGELICA HOLMES, University College London, United Kingdom

LU LEMANSKI, University College London, United Kingdom

MARIE VASEK, University College London, United Kingdom

ENRICO MARICONTI, University College London, United Kingdom

Location-based games have come to the forefront of popularity in casual and mobile gaming over the past six years. However, there is no hard data on crimes that these games enable, ranging from assault to cyberstalking to grooming. Given these potential harms, we conduct a risk assessment and quasi-experiment on the game features of location-based games. Using PokémonGO as a case study, we identify and establish cyber-enabled stalking as the main risk event where in-game features such as an innocent function to share in-game postcards can be exploited by malicious users. Users obtain postcards that are unique to each PokéStop and represent gifts that can be shared with in-game friends. The number of postcards that each user can retain is limited, so they send the excess to their friends with items that boost their friends' game activities. The postcard often also unintentionally leaks the users' commonly visited locations to their in-game friends. We analyze these in-game features using risk assessment and identify cyber-enabled stalking as one of the main threats. We further evaluate the feasibility of this crime through a quasi-experiment. Our results show that participants' routine locations such as home and work can be reliably re-identified within days from the first gift exchange. This exploitation of a previously unconsidered in-game feature enables physical stalking of previously unknown persons which can escalate into more serious crimes. Given current data protection legislation in Europe, further preventive measures are required by Niantic to protect pseudonymized users from being re-identified by in-game features and (potentially) stalked.

## 1 INTRODUCTION

In 2017, Nicci Kay came home from work to find a note left on her doorstep accusing her of cheating in Ingress, a popular mobile location-based game where players can interact with each other in proximity of their physical location [11]. Using the in-game features and information provided by unofficial third party tools, players were able to identify her home and stalk her. Further reports of misuse of location data emerged regarding other games such as PokémonGO, revealing that users were harassed and did not feel safe while playing given the numerous interactions with strangers in the public space [59].

The fusion of digital and physical spaces yielded by location-based games (LBGs) and augmented reality (AR) leads to the creation of hybrid spaces [12] where in-game features have repercussions in the physical world, similarly to what was experienced by Ms. Kay. In 2016 these games were revolutionized by PokémonGO (PoGO), a new game introduced by Niantic Inc. where players walk around in their world with the game's Pokémon world overlay. This allows users to catch their favorite Pokémon as well as interact with other PokémonGO players via trades, battles, and gifts. Although more than 6 years old, this game is still one of the most popular and most played LBGs. The popularity of the game and its features have caused several unexpected issues (e.g. players trespassing in military areas to capture a rare Pokémon) and, as a consequence, players put themselves and other members of the public at risk. Aside the players' lack of care and attention to what is around them in the physical realm, these novel in-game features could be exploited maliciously to lure players or follow them with criminal intents. We therefore ask: *can the in-game features of PokémonGO (PoGO) be used as a toolkit by offenders?*

We aim to identify the potential offenses facilitated through the in-game features of PoGO, discuss its related crime-prevention and policy implications and explore future avenues for research. To do so, we carry out a risk assessment and show that a variety of risk events can be perpetuated by PoGO 's in-game features. We identify 'cyber-enabled stalking' were offenders stalk victims in the real world enabled by location data leaked by the victims use of the app as the primary and most feasible risk event.

Given the lack of research on the malicious use of LBG and AR games to commit crimes and the explosion in their popularity, we consider and conceptualize the hidden risks within the structures of such games. Here, PoGO serves as a case study to identify the potential offending risk events enabled by its in-game features. Understanding how LBGs and AR features may operate in a criminogenic landscape is crucial to evaluate the security vulnerabilities of such applications. Our findings could be used to better inform players on how to protect themselves as well as allowing companies to design new games with the users' safety and security in mind.

Our work is carried out in two stages, Risk Assessment and Quasi-Experiment, after we review the literature in Section 2.

- Our risk assessment in Section 3 of PoGO identifies and screens the most relevant risks from an offender perspective.
- Our quasi-experiment tests and evaluates the feasibility of cyber-enabled stalking as the risk event in Section 4.
- We then discuss the implications of our findings in Section 5, both under the lens of safety, particularly because of the young age of PoGO users and under the lens of modern data protection legislation from the EU and UK.

## 2 BACKGROUND

### 2.1 The rise of LBGs and safety considerations

Geo-location-based videogames, also known as location-based games (LBGs), are games that use location-based technology such as GPS, QR codes or Bluetooth to direct users towards specific locations, where players can do specific tasks and interact with other players [12, 71]. Geocache, considered one of the first LBGs since its beginnings in the 2000s, uses GPS to hide containers all over the world for other players to use coordinates and find them. LBGs caught the attention of the gaming industry around the 2010s when smartphones allowed playing games while walking in public spaces. Developers began integrating into their gameplay augmented reality (AR), a technology that allows the creation of a digital environment that blends with the physical world, layering spaces beyond the virtual one [41].

Today, there are over hundreds of LBGs on the iOS and Android markets, the most popular being and having been Ingress Prime, PoGO, Zombies Run, Pikmin Bloom, The Witcher: Monster Slayer, Jurassic World Alive and Harry Potter: Wizards Unite. The first safety concerns emerged with Ingress, when inadvertently, players began attracting law enforcement attention because of what was deemed as suspicious behaviour [23]. Further reports included game features directing players to unsafe places or facilitating stalking through location base data [11, 73].

Launched in 2016 by Niantic, Inc, the LBG PokémonGO or PoGO, raised players' safety concerns which have been the forefront of academic discussions. Despite the media panic around this game [45], its location-based technology carries legitimate risks for users. Studies have demonstrated the increase in unsafe driving or walking around traffic, leading to increased vehicle crashes and incidents caused by inattention while playing [16, 29]. Playing in extreme weather conditions or in dangerous locations also poses a risk to personal safety [47]. PoGO has led to numerous cases of trespassing, as players violate the private properties and privacy of individual citizens in their homes but also of public spaces such as museums, places of religious worship or cemeteries. Mukhra et al. [50] explore the ethical concerns around PoGO but go beyond a description of 'risks as hazards' posed by the game by analysing its relationship with crime. PoGO has led to new opportunities for software hacking, violent robbery [22] or homicide. Here, the nature of the risk is not environmental but relational as players encounter risk posed by other players [73]. Players become vulnerable to crime from the juxtaposition of the places and players they meet. Awareness of this can be seen in player's reluctance to walk in certain areas because they do not feel safe or at risk of violence [64, 67].

Privacy concerns are a related concern, particularly concerning the privacy of minors on mobile apps. A number of authors have done wide-scale investigations of privacy on apps mobile apps [2, 61, 77]. This work generally focuses on technical features inside the apps that can be directly inferred through techniques like reverse engineering. Harbouth and Pape surveyed users about privacy concerns in augmented reality apps [31]. Harbouth and Frik then evaluated the privacy permissions models in these apps and proposed ways of more clearly communicating more complex risks stemming from the real world dimension to users [30]. While they consider, among other factors, privacy risks of users working together on an augmented reality object, most of the concerns raised here were regarding metadata collected by a phone and then passed on to others. This work also relates to work surrounding location-based privacy via geotags [69, 72].

## 2.2 Crime Science Theories and their applicability to LBGs

Crime Science is a data-driven and scientific approach to analyzing factors that create opportunities to perpetrate crime. Routine Activity Approach (RAA) and crime pattern theory posit that crime is not random and occurs in specific spatio-temporal patterns [6, 18, 75]. Offenders will commit their crimes in locations they are familiar with and will often be part of their daily life. This means that if a potential offender can easily identify a victim's daily routine, they could frequent the same locations, becoming more familiar with the areas and therefore more comfortable and willing to commit a criminal offense.

As 'crime feeds ordinary routines' [17], it is worth noting the potential for LBGs and specifically PoGO to reveal player's routines. The fundamental idea behind these games is in fact to intertwine LBGs and AR technology with player's everyday life, enhancing daily experiences through inter-active gameplay [33, 51]. PoGO is integrated in the routine of its users as most players play on their way to work, to run daily errands or during their leisure time. Specifically, what is important for players is the ability to play while 'doing something they were going to do anyway' [43], exemplifying the superposition of PoGO play with users' routines in space and time. According to the geometry of crime theory [4, 5], the routines and movements of victims are similar to those of offenders: crime happens where and when both intersect.

With more than half of players playing several times a day [55], the opportunities for crime are plenty. This is especially relevant to young players, who are more likely to have a cyclical and strict routine. Lack of safety was a main theme of concern for parents [46]: when co-playing with their children, they reported playing mainly in their neighborhood or at nearby parks they considered 'safe' [67]. In a qualitative examination of Pokémon card and children's geographies, Horton [35] finds that participants' space-time routines were even altered to accommodate play, translating into physical detours to shops and hangouts (despite going against the advice of parents and teachers). This is an example of how LGBs mould users' daily routines and experiences of space which is significant to conceptualise its relationship with crime. The superimposition of crime science theories with empirical evidence of their applicability to games such as PoGO highlight the urgency to think about the crime opportunities inadvertently yielded by this new technology.

## 2.3 Risk, Risk Assessment, and Crime

The definition of the term risk varies across the literature and has seen a variety of definitions proposed.

(1) Risk is the 'combination of the probability of an event and its consequences' [20].
(2) Risk is the uncertain consequence of an event or an action on something of value. Risk in this definition is therefore the combination of probability and severity of consequences linked to natural events, human activities, or any combination of the two [54].
(3) Risk is the effect of exposure to uncertainty on objectives relevant to an individual or organization [34, 40].
(4) Risk is the combination of events, consequences, and the associated uncertainties [3].

From these definitions, risk is generally defined with respect to the consequences associated with an uncertain event. The probability of an event occurrence and the effects (both positive or negative) of the consequences can be used to characterize types of risk [19]. Exposure to risk can result in varying consequences that may either bring on positive effects (opportunities), negative effects (threats), or any combination of both [38]. It is therefore important to manage risk to optimize the opportunities afforded by positive risk, while avoiding or mitigating negative risks or threats that may result in problems that harm the individual, organisation, or system [19, 38]. This can be achieved through the application of risk management frameworks.

Much like the variety of definitions of risk, there equally if not more standards and frameworks to enable the management of risk. The COSO Enterprise Risk Management-Integrated Framework (ERM) and ISO31000 standards are two of the most well-known and widely implemented risk management frameworks [24, 25]. While the COSO framework and ISO31000 supports risk management through the emphasis of different aspects of individual and organizational goals such as regulatory compliance (COSO) and management processes (ISO31000) [27], both internationally implemented standards show that risk assessment is a key component which continuously interrelates with its surrounding principles and components [24]. In both standards, risk assessment is divided into three core procedures: risk identification, analysis, and evaluation [40, 49].

Risk management and assessment can be extended beyond the enterprise context. Risk assessment has been a tool in the criminological context, specifically in relation to sentencing, policing, and crime prevention [66]. Risk assessment in the criminal justice context retains the three core steps of identification, analysis, and evaluation. The SARA (scanning, analyzing, responding, assessing) model in problem-oriented policing enables practitioners to identify, analyze, respond, and assess a problem (risk) in a cyclical manner [8]. Crime Prevention Through Environmental Design (CPTED) is a rational risk management approach that seeks to prevent crime and security risk by identifying, analyzing, and evaluating potential vulnerabilities in built environments or objects of interest/value [40]. Further, CPTED not only complies with all iterations of the ISO31000 standard [10], it has since been incorporated as the key underpinning concept and framework behind ISO22341, the international standard for security and resilience [21].

## 3 RISK ASSESSMENT

We carry out a risk assessment on the potential of risk events through the exploitation of LBG features in PoGO using the guidance provided in the risk management process found in the ISO31000:2018 standard [40]. Definitions of commonly used terms from PoGO can be found in Appendix B.

### 3.1 Risk identification: Potential offences facilitated through PoGO

Finding, identifying, describing, and recording risks that could aid or obstruct the system or organization are the main goals of risk identification [40, 49]. Events, circumstances, or situations which could have an impact should be identified in this process [39, 40]. In addition to the risk event, information on their cause and possible effect/consequences should be considered and recorded in this step. While there is a wide array of risk identification tools and techniques, there is no single 'best method', rather, an appropriate selection and combination of approaches should be used [32, 65]. Using the information provided by ISO/IEC 31010, we select preliminary hazard analysis (PHA) and brainstorming as the combination of risk identification approaches. Through PHA, a list of risk events is generated and the potential points of failure and consequences are further identified by brainstorming. Both PHA and brainstorming satisfy ISO/IEC 31010 as they are strongly suited for risk identification with novel technologies or circumstances when data is limited [39].

Using this process, we identify 11 risk events (ie. criminal offenses) that may be facilitated through the main in-game features of PoGO (Table 1). We divide them into two categories: *opportunistic* events where the offender does not go into the situation intending on committing a crime and ends up committing one and *premeditated* events where the offender plans on committing a crime or crimes (regardless of whether the offender has identified a victim or not). These are: theft/robbery (opportunistic or premeditated), assault (opportunistic or premeditated), manslaughter, drug dealing, cyber-stalking, cyber-enabled stalking, murder, grooming and (premeditated) sexual assault. We refer to in-game features as the main mechanisms that use GPS location/AR technology within the game to connect, play or interact between users. These include: Pokéstops and gyms, Gift exchange, Trading, Trainer battles, Raids, Friends list and Lures. A detailed definition of each in-game feature is listed in Table 4 located in Appendix A and Figures 1, 2, 3, 4, and 5 which are in-game screenshots of these features.

### 3.2 Risk analysis

*3.2.1 Cyber-enabled stalking as a high probability risk event.* The second stage of risk assessment is risk analysis to understand the potential causes and consequences of the risks in the context of the current system. This enables us to establish the full picture of the risk event, point to relevant, impactful weaknesses, that will hopefully point to risk mitigations. Risk assessment techniques can be qualitative, quantitative, or mixed.

After identifying 11 potential risks in Section 3.1 as shown in Table 1, we screen this list to identify the risk event that warrants more detailed assessment. We consider the consequences ("strongly applicable"), probability of event ("strongly applicable"), and level of risk ("applicable") individually in Appendix A using Consequence-analysis. By considering the risk event as the consequence, we take into consideration the contributory factors and interactions of PoGO in-game features and the secondary consequences that could arise after the risk event has occurred [39].

Table 1.  Identified risk events and the associated in-game features in Pokémon Go.

| Identified Risk Events | In-game Features | | | | | | |
|---|---|---|---|---|---|---|---|
| | Gift-exchange | PokéStops & Gyms | Trading | Trainer Battles | Raids | Friends List | Lures |
| *Opportunistic Events* | | | | | | | |
| 1. Theft/robbery | | ✓ | | | ✓ | ✓ | |
| 2. Assault/sexual assault | | ✓ | | | ✓ | ✓ | |
| 3. Manslaughter | | ✓ | | | ✓ | | |
| *Premeditated Events* | | | | | | | |
| 4. Theft/Robbery | | ✓ | | | ✓ | ✓ | |
| 5. Assault/ Kidnapping | | ✓ | | | ✓ | ✓ | |
| 6. Drug Dealing | | ✓ | | | | ✓ | ✓ |
| 7. Cyberstalking | | ✓ | ✓ | | | ✓ | ✓ |
| 8. Cyber-enabled stalking | ✓ | ✓ | | | | ✓ | |
| 9. Murder | ✓ | ✓ | | | | ✓ | |
| 10. Grooming | ✓ | ✓ | | | | ✓ | |
| 11. Sexual assault/ rape | ✓ | ✓ | | | | ✓ | |

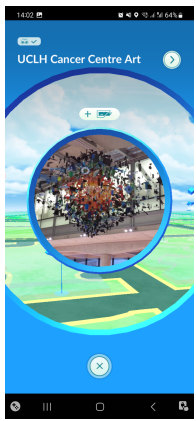

Fig. 1.  In-game screenshot of the stops

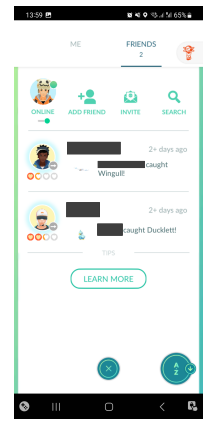

Fig. 2.  In-game screenshot of the friends list

We consider how the offense could be committed before assessing the severity of the consequences and estimating the attack probability. For example, while assault and drug dealing show high severity of consequences, we consider the probability of them being carried out to be 'low' due to the cost and amount of effort posed by each scenario. While the idea of using Pokémons in Trainer battles to represent drugs could be appealing, its lack of practicality makes it an unlikely scenario to be adopted by offenders. Now, the conceptualization and assessment of each scenario is hypothetical, but this does not mean it has never occurred. The lack of reliable empirical evidence justifies the necessity to systematically examine the possible risks events that could be carried out.

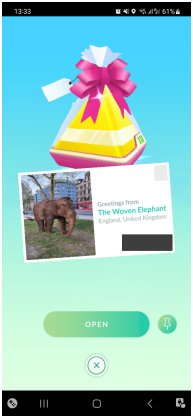

Fig. 3. In-game screenshot of the gifts

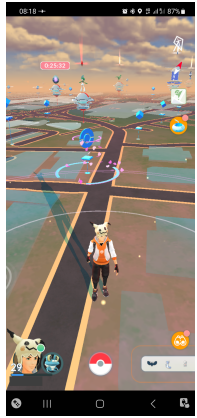

Fig. 4. In-game screenshot of an active lure on a PokéStop

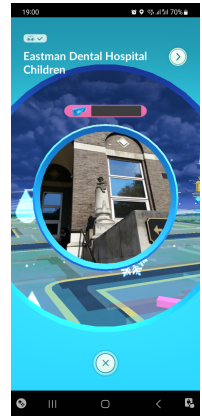

Fig. 5. In-game screenshot of the user activating the lure at the PokéStop

We establish that cyber-enabled stalking as the highest probability risk event for two main reasons. First, the severity of its **consequences** is high. Victims of cyber and physical stalking incur both physiological and psychological consequences involving mental and emotional distress (prolonged worry, chronic low mood, fear, irritability, helplessness, depression, anxiety), cognitive issues (poor concentration), and development of a suspicious nature or even PTSD, which can result in self-isolation [44]. However, looking beyond the specific consequences of cyber-enabled stalking, we find that this risk event can act as the basis for other risk events. Appendix A shows that risk events such as murder, grooming, and sexual assault use the same in-game features as cyber-enabled stalking.

Second, the **probability** of this risk event is high. The exchange of gifts with postcards containing identifying names of the PokéStops most visited by the player gives the offender a tangible and identifiable spatio-temporal routine of his target victim. This routine can be easily visualised by inputting the names of PokéStops or landmarks directly on a map (such as Pogomap or Googlemap, respectively Figures 7 and 8). The effort to carry out this risk event is relatively low as it uses a core game play feature of gift exchanges and does not require offenders to stand at specific PokéStops location unlike theft/robbery or drug dealing. But what is crucial here is that cyber-enabled stalking has already been reported to exist in PoGO, albeit from anecdotal sources [36, 42, 59, 63]. Without serious examination of the matter, there is no evidence that researchers and game developers can use to understand and prevent this risk from happening.

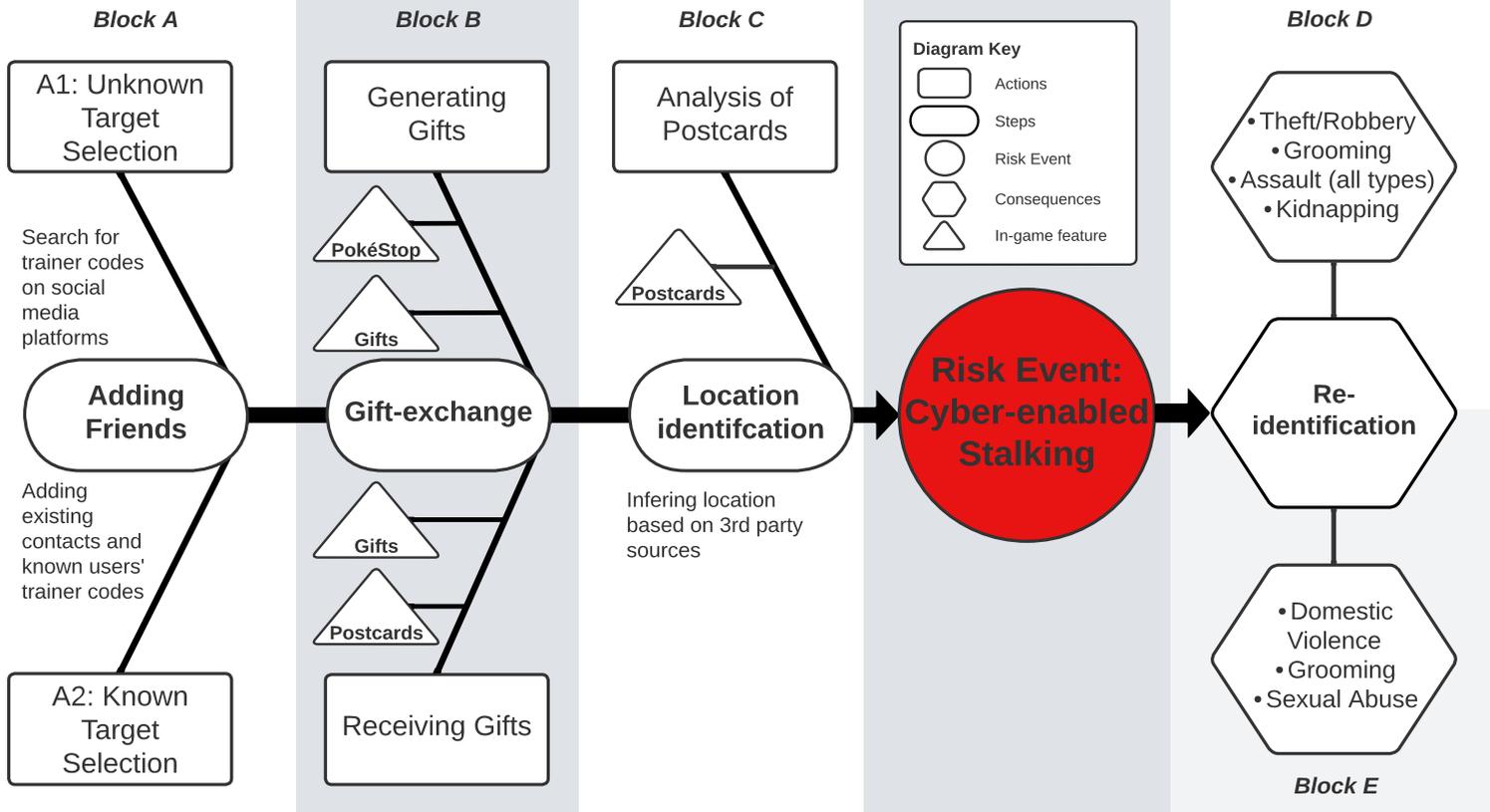

Fig. 6. Modified Ishikawa diagram showing the causes and the sub-causes that lead to the risk event (cyber-enabled stalking). This diagram also shows the consequences of the risk event.

*3.2.2 Conceptualizing causes and consequences of cyber-enabled stalking.* Having established cyber-enabled stalking as the risk event for further investigation, we selected a cause-and-effect analysis to identify the possible causes of cyber-enabled stalking. This method of analysis enables us to identify possible root causes, contributory factors and the interactions of the in-game features in PoGO that could lead to cyber-enabled stalking, thereby providing a basis to develop a potential hypothesis for further study. Using a modified Ishikawa diagram (Figure 6), we detail how offenders can exploit PoGO in-game features which lead to cyber-enabled stalking and other offences and show the consequences of the risk event. Throughout this process, we refer to the block letters or steps in the diagram.

First, offenders need to add their target as a in-game friend (Block A). There are two different possibilities here: the offender is either a stranger to their target (Block A1) or a known acquaintance (Block A2). In the stranger scenario, websites or social media platforms are used to find their trainer code which allows them to add them as a friend. Conversely, acquaintances may already have their target's trainer code or can add each other by importing or inviting existing contacts. Offenders then need to engage in the process of reciprocal gift exchange (Block B). This consists of 'swiping' PokéStops to generate gifts and sending them to the selected friend which needs to reciprocate the action and send back a gift to the offender. By sending the gift, the receiving user will be able to see the location from where that gift was generated from. The location comes in the form of a postcard that can be saved. Therefore, the more gifts exchanged, the more location information is available to the offender. The repetition of this step leads to the analysis of the postcards to identify the general PokéStops locations from which the gifts were sent (Block C). Locations and routines can be inferred from the frequency of receiving a gift and by entering the name of the PokéStops into third party sources to locate the stop. For example, a user can search for the PokéStop called 'New York Public Library – Batt' on Pogomap.info and be shown the exact location and coordinates of that location (Figure 7). Third party sources not specific to PoGO such as Google Maps will also make the correct location deduction most of the time as users are able to identify the PokéStop location using the same search string (Figure 8).

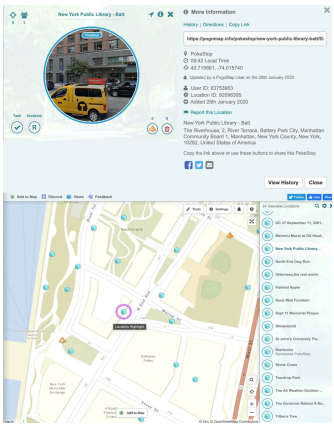

Fig. 7. PoGo map representation

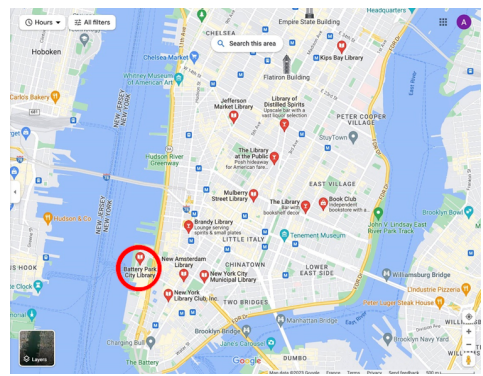

Fig. 8. Google maps representation

The completion of these three steps enables the risk event, cyber-enabled stalking, to occur. The immediate consequence of cyber-enabled stalking is the re-identification of the player. We refer to re-identification here broadly as "any process that re-establishes the relationship between data and the subject to which the data refer" [26, 28]. This means that a player whose real identity has been pseudonymized from the game through usernames and unique trainer codes, can be re-identified by cyber-enabled stalking.

Thus, the target player can be re-identified in the physical world and approached by the offender. This is a significant consequence brought by the risk event because it is the necessary cause for carrying out other types of offenses. In the stranger scenario, first time offenses such as assault, kidnapping, theft/robbery or grooming may be more likely to happen (Block D), whereas the acquaintance scenario may be more likely to lead to the re-victimisation of the target (Block E). Offenses such as domestic violence, sexual abuse or grooming may be more characteristic of repeat victimization. However, we acknowledge that crime types are not mutually exclusive to each scenario; we hypothesize that these types of crimes may be more prevalent within these cases.

## 4 QUASI-EXPERIMENT

### 4.1 Rationale: from conceptualization to practice

Section 3 establishes cyber-enabled stalking as a high probability risk event that acts as an enabler for other offenses since the identification of the target victim is possible through the in-game features of PoGO. However, there is no empirical data to prove the feasibility of such process. We therefore conduct a quasi-experiment to test whether the in-game features of PoGO can be used to identify the routine locations of the target user. We base the quasi-experiment on the 'Stranger Scenario' (Block A2 shown in Figure 6), since the main obstacle to initiate the cyber-enabled stalking risk event is for an offender to add a potential target user to their friends list in PoGO. Using this rationale, we form the following hypotheses:

$H_1$: *It is possible for a potential stalker to identify the PokéStops closest to a user's home and work within 1 kilometre (0.6 miles) range through the exchange of gifts on PoGO.*

$H_2$: *The increase of gift exchanges on PoGO will allow for more accurate determinations of the PokéStops closest to a user's home and work.*

To test the hypotheses, we use a multi-stage design to ensure external validity and quality data collection constrained by the time frame and cost restrictions of the study. We aim to establish a proof-of-concept regarding the feasibility of location identification, therefore there is no need for a control or pre-test group prior data collection [74]. The experiment is split into two stages: survey phase and gift-exchange phase, each phase was carried out individually by Researchers 1 and 2 respectively (Fig.9). This procedure ensures internal validity as the location guesses done by Researcher 2 are not influenced by the information provided in the surveys.

### 4.2 Data Collection: Survey Phase

We developed an online survey to gather information on participants' regularly frequented locations such as the three closest PokéStops to their workplace, leisure, and home, gaming habits with regards to PoGO, trainer code, how many friends they had in PoGO, and whether they knew them personally outside of PoGO (full text in Appendix C). To fully mimic the how an offender might seek out potential targets in the 'stranger scenario' (Block A2 from Figure 6), Researcher 1 recruited participants in gaming communities and groups on key social media platforms such as Reddit and Discord. The eligibility criteria further required participants to be over the age of 18, and London-based. We have chosen London as it is a large metropolitan city where most of the population uses public transport for their activities and tends to live in outer areas and travel to the city centre for work. Thanks to the characteristics of the city, the experiment reflects real world decisions where offenders prefer areas of familiarity based on the theories posited by the routine activity approach [17].

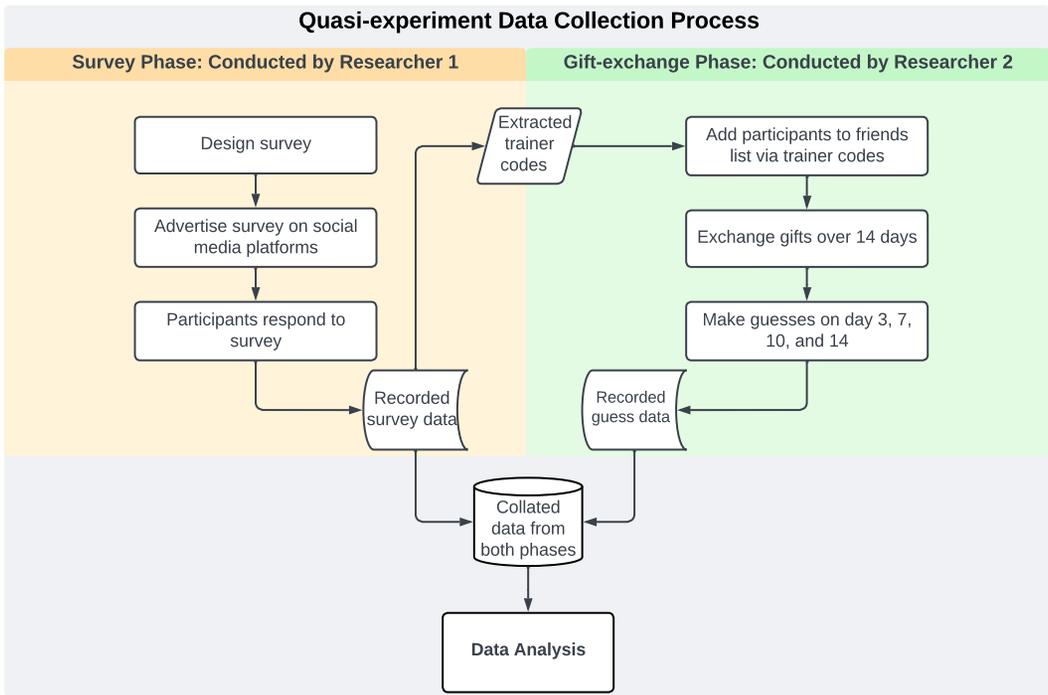

Fig. 9. An overview of the data collection process, split into two distinct phases. Each phase was carried out by Researchers 1 and 2 respectively.

### 4.3 Data Collection: Gift-exchange Phase

After the survey data collection run by Researcher 1, Researcher 2 added each participant in PoGO via the friends list feature using the trainer code provided and sought to exchange gifts with them daily. The experiment ran for two weeks – Mcewan et al. [48] stated that that this duration effectively differentiates between brief harassment and problematic stalking behavior.

In the gift-exchange phase, Researcher 2 records the name and location information from the postcard included in the gift received. We also record screenshots of each gift received for verification but, once we double-checked the correct work, we deleted them for privacy preservation. We use the PokéStop map (pogomap.info), as shown in Section 3.2.2 to aid our guesses on the PokéStop location which generated the gift exchanged, adhering to the sub-causes of cyber-enabled stalking in Block C of Figure 6. Given that users use the app throughout the day, Researcher 2 makes guesses about the three closest PokéStops to each participants' workplace, home, and leisure locations along with an overall confidence rating for these. Researcher 2 guesses the locations on days 3, 7, 10, and 14, but none of these guesses are confirmed nor denied until this phase was completed.

### 4.4 Ethics

This research was deemed exempt from formal ethics review by our department's ethics committee. The in-game features have not changed since the experiments were run. We are in the process of responsible disclosure towards Niantic that will be concluded before the publication of this work. The participants involved in the experiments were fully informed (see supplementary material) and signed a consent sheet but the structure of the questionnaire allowed them to preserve at least pseudoanonimity throughout the whole experiment. To mitigate the risk of localising the participants in case of a rogue researcher we did not ask the exact position but the three closest PokéStops. Participants were additionally asked to provide up to three leisure locations that they regularly visited. As these would not be considered private or identifying locations, it was not specified to provide PokéStop information but left for the participant to decide how specific to be.

## 4.5 Data Analysis: Methodology

| Accuracy of PokéStop guesses | Points attributed |
|---|---|
| Exact guess | 3 |
| Guess within ½ a kilometer of a stop the participant named | 2 |
| Guess between ½ and 1 kilometer of a stop the participant named | 1 |
| Guess more than 1 kilometer of a stop the participant named | 0 |

Table 2. Point system for location guesses during the gift-exchange phase.

We calculate the accuracy of guesses using a point system (Table 2) and then convert to a percentage for standardization. Any guess over 1 kilometer is considered unsuccessful given the inability for a stalker to identify their target. We asked the participants for three of the closest PokéStops to their home and work, allowing for a possible total of 9 points each. The points scored by Researcher 2's guesses will then be presented as a percentage of the 9 points maximum score. Leisure locations were inconsistent as some participants did not provide any PokéStop locations.

We use data from the previously collected survey and gift-collection phases to determine how quickly and accurately an individual's frequently visited location could be determined via PoGO gift exchange as well as how the number of gifts exchanged affects the accuracy and confidence of these guesses. Additionally, we examine data gleaned from habits of participants to determine whether risk events like cyber-enabled stalking are feasible by potential offenders.

## 4.6 Data Analysis: Results

A total of five participants responded to the survey, of which only 4 participants took part in the gift-exchange phase of the project. This small sample size can be attributed to time restrictions and difficulties surrounding advertising in private groups. Nevertheless, the participants provided a varied dataset, representing different age groups, play times, and commitment to participation, allowing for comparisons between them.

Figure 10 summarises our findings. Participants 1 and 4 both sent gifts every day during the gift-exchange phase. Participant 3 sent gifts almost every day, missing only two days throughout the two-week period. Participant 2 only sent two gifts within the first week of gift-exchange phase.

We calculated the accuracy percentage as explained in Section 4.5 and analysed the results in relation to the confidence Researcher 2 had when guessing the locations. There was no improvement in the home location accuracy guesses throughout the two-week period for any of the participants. Researcher 2 achieved the highest accuracy with 88% and 77% for Participants 4 and 1 respectively while guesses for Participant 2 only attained 33%. Although guesses were made for leisure locations, there were no correct guesses for any participants throughout the two-week period, therefore the visualization of this data has been excluded. We now study each participant individually to see how each person's behavior adds to our understanding of their behavior, which we will later see through the lens of cyber-enabled stalking.

*4.6.1 Participant 1.* For the home location, one PokéStop was correctly guessed, and the two others guessed were within half a km of the correct location. The three guesses remained the same from day 3 through 14 for home guesses leading to no change in accuracy. The participant provided the same location for work and home. Researcher 2 initially guessed that the participant worked from home or was unemployed in the day 3 and day 7 guesses, leading to the same accuracy score as for the home location. With increasing information from gifts, Researcher 2 made different location guesses on days 10 and 14 causing the accuracy score to drop to 0% in the second week of guesses. Guesses were made for leisure locations; however, the participant did not provide any information on leisure locations they frequent.

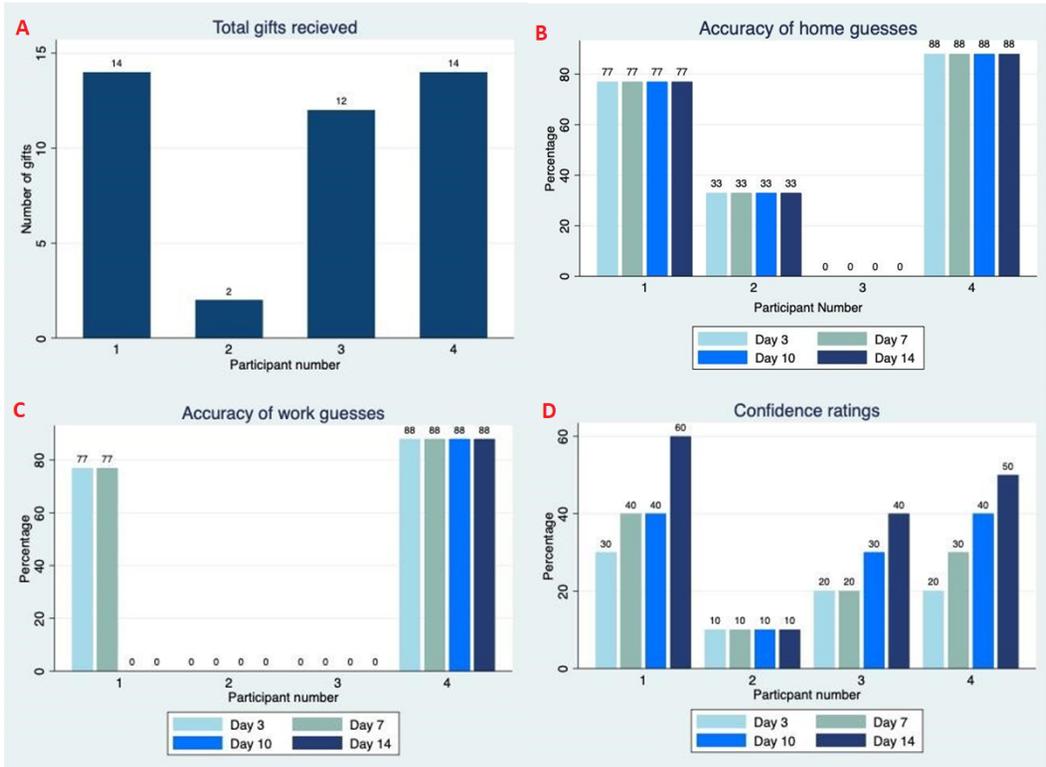

Fig. 10. A: Total number of gifts received from each participant. B: Accuracy of guesses relating to home location. C: Accuracy of guesses relating to work location. D: Confidence rating on each guess day.

Some of the locations from which gifts were sent could not be found on the searchable PokéStop map. Many locations, especially remote ones such as those outside main cities and hiking trail markers, had not been added to the map as of the time of the study. Although the exact location of the PokéStop could not be found, it was always still possible to identify a general area given the information included in the gift.

*4.6.2 Participant 2.* Participant 2 only sent two gifts, both within the first week of the in-game data collection. This lack of gifts had a significant impact on Researcher 2's confidence ratings as there were no additional data points to base guesses on. Although the confidence ratings had no impact on the accuracy score, they were used in this study as a pseudo-gauge of how confident a potential stalker would be in their estimation of an individuals' location, and therefore how likely they might be to physically attend the location to find the person. In the case of this participant with such a low confidence rating, it might be assumed that a stalker would not feel comfortable or confident enough to visit the location or might simply not want to waste their time doing so. Despite this, one PokéStop was correctly identified for the home location. Only one other PokéStop was guessed by Researcher 2, however, this was more than a km from any of the correct PokéStops provided by the participant. However, this guess, although not within the required distance, was within the correct town. This participant did provide separate locations for work and leisure both of which the researcher failed to guess.

*4.6.3   Participant 3.* This participant only failed to send gifts for 2 days during the first week, day 2 and day 5, explaining a lack of change in confidence rating from the researcher from the day 3 guess to the day 7. With daily gifts in the second week, the confidence ratings increased. Researcher 2 achieved the lowest accuracy scores despite a consistent number of gifts received. If parameters were expanded to allow for any correct guesses in spite of the location category the researcher believed them to fall under, Researcher 2 would have achieved 66% in accuracy, as guesses made for the leisure locations were all within $\frac{1}{2}$ km of the home PokéStops provided by the participant. Here the researcher incorrectly assumed that gifts from central locations are more likely near work and further away are more likely near home. We received all this participants' gifts from PokéStops close to Participant 3's home and no gifts from locations close to their work or leisure locations.

*4.6.4   Participant 4.* For the home location, Researcher 2 guessed two PokéStops correctly and one within $\frac{1}{2}$ km with an accuracy of 88% on every guess. Similarly, the work location had two correct PokéStops and one within $\frac{1}{2}$ km. Guesses made on day 3 remained the same until day 14, giving a consistent accuracy score. As with the other participants, there were no correct guesses for the leisure locations. The researcher did not make any guesses on leisure for days 3, 7, and 10. On day 14, the researcher did guess one PokéStop for leisure. The gift received informing this guess was one from which was not on the PokéStop map and was close to the participant's work location which the researcher did not guess. The confidence rating consistently increased with each guess made for this participant, however it never reached as high as for participant 1 due to more difficulties in finding the exact PokéStops on the map being used. This participant was the only one who did not play every day as reported on our survey.

## 5   DISCUSSION

The results of the quasi-experiment demonstrate not only the feasibility, but the effectiveness of cyber-enabled stalking using the gift exchange feature through PoGO. It was possible to guess the home and work locations for 2 out of the 4 participants to within $\frac{1}{2}$ a kilometer. The inability to accurately guess locations for 2 participants stemmed from the lack of gifts (Participant 2) sent and the assumption made by the researchers that the work place would be located in a more central area than the participants' home (Participant 3). With working from home becoming far more prevalent particularly since 2020 [14], the home and work locations were more likely to merge. While useful for this experiment, this categorisation of locations may not be relevant when applied to a real-life scenario. Should a stalker be using PoGO as a tool to identify their victim's location, it is unlikely they would feel the need to make such distinctions as it would be of little importance if the location was work or home if it was frequently visited. The stalker could visit the location physically and identify environment what relationship the individual had with said location. Given the results of the quasi-experiment, we confirm hypothesis 1. It is empirically possible for a potential stalker to identify the PokéStops closest to a active user's home and work within (in the worst case) 1 km (0.6 miles) range through the exchange of gifts on PoGO.

The second hypothesis stated that increasing gift exchanges between users would allow for more accurate guesses. During the experiment, the only changes in accuracy occurred at day 10 with participant 1 for work location guesses, drastically dropping from 77% to 0%. Given the limited sample size, no relationship can be confirmed. However, this suggests that with more data points, accuracy of guesses may in fact decrease as the guesser becomes confused by the introduction of new locations. This echoes the apparent lack of relationship between confidence ratings and accuracy scores. Therefore, we cannot confirm hypothesis 2.

We find that the map used to identify PokéStops did not have all of them marked, especially when looking at more residential and rural areas outside of the city. These could not be searched and found on the map using only the name of the PokéStop. It was still possible to identify a general area. Combined with information from previous postcards received, this could enable a higher accuracy guess.

## 5.1 Implications

The implications of our findings are varied. We first discuss the immediate consequences on safety and show repercussions on data privacy.

### 5.1.1 Safety Implications.

The ability for PoGO and potentially other LBGs to overlay and reveal the daily routines of users poses considerable safety implications given how these patterns can be exploited. In crime pattern theory, key locations such as home or work are referred to as nodes and the routes between them as paths [75]. Here, PokéStops act as nodes that can proxy the location of a user's home or work. This enables offenders to not only identify locations and activity spaces where a target is likely to be, but further infer the relationships and paths between them.

It is often forgotten that offenders too have spatio-temporal routines and tend to behave like the non-offending population [6]. Frequently visited PokéStops or Gyms as part of an offender's own routine can intersect with those of the victim. Given the relative high accuracy gained from location information in the first gift exchanges, stalkers may be more willing to spend more time or walk further to a PokéStop based solely on one gift, regardless of their confidence [62]. The core tenet of routine activity theory is that a motivated offender and a suitable target must intersect in time and space without the presence of a capable guardian, manager, or handler [8]. The in-game features identified in this risk assessment and tested in the quasi-experiment show that motivated offenders can gain location information and 'forcibly' intersect with the user. Cyber-enabled stalking shows that it is the *cyber*/AR space that first allows the identification of the physical space. Being 'in the same place at the same time' is no longer the issue. The intersection of targets' and offenders' spaces in PoGO may increases the risk of offences because the 'journey to crime' [60] is facilitated: seeing the locations where a target is likely to be may increase the distance a stalker is willing to travel to commit an offense.

Cyber-enabled stalking poses more serious safety implications for PoGO users as it can serve as a basis for further crimes. It has long been recognized that cyberstalking is often accompanied by or develops into more traditional, real-world behaviors. The majority of the cyberstalkers also engage in offline forms of stalking, demonstrating the potential for online offenders to develop or escalate their behavioral repertoire of stalking or violent behaviors [7]. Despite less frequently reporting fear, reported threats and physical attacks were higher for cyberstalking victims compared to those stalked offline only [53]. The risk analysis identified subsequent offenses as consequences enabled by cyber-stalking such as assault, grooming or domestic violence (Fig 6). There is evidence that cyberstalkers are more likely to have been intimate partners with their targets, compared to traditional stalkers, and were not more likely to target strangers [7]. However, the data collection period for this study (2002-2013) preceded the advent of PokémonGO and AR gaming by several years. It is likely that the creation of new opportunities to monitor strangers' locations could have resulted in more activities targeting unknown individuals that could result in cyberstalking progressing to real-world interactions.

The characteristics of PoGO players may make them more likely to become offenders and victims of cyber-enabled stalking. Research indicated cyberstalkers tended to be male, younger, educated, well-performing, and technologically savvy, compared to traditional stalkers [44], which, given the wide PoGO player-base, can include individuals with those characteristics. Increasing time spent online is positively correlated with engaging in cyberstalking [68]. Likewise, levels of engagement with cyberspace (e.g. degree of disclosure online) can impact an individual's experience of cyberstalking, as it is suggested that frequent and prolonged Internet use can increase the likelihood of being targets [44]. This raises serious implications for the PoGO playerbase who are prone to spend increasing amounts of time online.

### 5.1.2 Data Privacy Implications.

The in-game features we identified in PoGO that may be associated with cyber-enabled stalking are not novel when considering the wider LBGs landscape. LBGs seek to incorporate physical world locations as a core gameplay feature to encourage user interaction. However, this raises privacy related concerns in addition to the safety implications of the risk assessment and quasi-experiment conducted.

Within the definitions and scope of the EU General Data Protection Regulation (GDPR) and the UK Data Protection Act 2018 (DPA), personal data pertains to any information relating to an identifiable living individual [15, 70]. Information such as - name, identification numbers, or location data, are considered identifiers. Using the definition provided by the National Institute of Standards and Technology, de-identification is a process by which information that contains identifiers specific to individuals is removed through a collection of approaches, tools, and methods [26]. It is also noted by NIST (IR 8053) that partial de-identification can lead to more information retrieved by the offenders. In our experiments, this means that the opportunity given by the game features of de-identifying the players location opens to more data being *re*-identified by a cyber-enabled stalker.

Location data and LBGs are intrinsically intertwined – LBG developers such as Niantic clearly state in their Privacy Policy and Terms of Service that usage of Personal Data such as location information and other interactions are necessary to provide the desired gameplay/services promised[1].These terms and conditions are agreed by users when the Niantic services such as PoGO are used. As a part of their role of as the data controller for PoGO, Niantic has determined the appropriate and acceptable trade-offs between de-identification and gameplay. Niantic has done their due diligence when examining their approach to PoGO in the context of the GDPR checklist [37], ensuring de-identification of their users through replacing usernames with trainer codes and maintaining pseudonymity by not using personal information such as date-of-birth, real names, gender where possible. In addition, PoGO outlines their Child Online Privacy Protection Act compliance, meaning parental consent must be given for personal data to be collected and processed.

Given the findings of the risk assessment and subsequent quasi-experiment, we question whether Niantic's measures of de-identification are enough to protect users from potential offenders. As we have identified in the risk assessment, sub-causes of cyber-enabled stalking utilize location data in postcards during gift exchange to eventually establish the routine of a potential target (Figure 6). It is stated in Clause 6 of Niantic's Terms and Conditions that collection and/or storage of any personally identifiable information of other users from Niantic's services without permission is prohibited. There is no feasible way for Niantic to detect this behavior; the collection of the target's location information can be done offline, therein bypassing the alert notification that a target would otherwise receive if their postcard had been saved by the offender in-game.

---

[1]https://nianticlabs.com/terms

The UK Data Protection Act 2018, s 171(1) states that reckless re-identification of de-identified personal data without the consent of the controller is an offense [70]. Using cyber-enabled stalking as the example, the 'de-identified personal data' would be the location data provided by the postcards during gift exchange. This act of re-identification through postcards can be seen as a violation of the DPA. However, sub-section (7)(b)(i) shows that the offender can be defended if consent was given by the data controller to process the 'de-identified personal data.' With this sub-section in mind, the following questions arise when considering the criminality of cyber-enabled stalking:

- What is consent with regard to using location data as a core part of gameplay services?
- Is consent for the processing of personally identifiable information of other users such as location data from postcards in gift exchange implied by virtue of users engaging in gift-exchange?
- Is consent for the processing of personally identifiable information of other users such as location data from postcards in gift exchange implied by virtue of PoGO – the data controller enabling this game feature?
- Should consent for obvious location-leaking behavior (like GPS tracking) and indirect location-leaking behavior (like gift exchange) be treated similarly?

While stalking is unquestionably a criminal offense, the 'cyber-enabled' element of this risk event could be defended if answers to the questions above remain ambiguous. Although the scope of Niantic's Terms of Service would result in termination of offender accounts if they were found to have violated the specified terms regarding processing personally identifiable data; this behavior is virtually undetectable until physical stalking has manifested. Given these implications surrounding GDPR and DPA, and the potential consequences of cyber-enabled stalking, it is perhaps pertinent to consider whether Niantic should be held to a greater degree of responsibility as the data controller to protect their player base from such risks. Perhaps PoGO and similar LBGs that use location data as a core gameplay feature could consider more rigorous measures to de-identify location information.

## 5.2 Considerations and Future Directions

Given the possibility for criminal offenses to be facilitated through PoGO, players can adopt various strategies to avoid being victimized such as not using real names, practicing caution prior to disclosing their information online [56] and refrain from playing in places or at times that may be dangerous. However, bearing the responsibility of safety on users assumes an understanding of online safety and how these translate to in-game features. Players may not know that trainer codes can be changed, how to report a player or service within the platform or how to access the Safety and Security Q&A in-game. Minimizing the risks for criminal offenses when playing a game should not be the responsibility of the user, but that of the game developer. The lack of digital literacy for players (and parents) may result in a lack of awareness of existing security measures or resources. It is therefore essential that platform providers consider these security risks as well as implement appropriate broad-acting safeguards that allow everybody to play the game safely.

While platform providers have an ethical responsibility to encourage all users to adhere to appropriate norms, there is also some discussion concerning their access to information which could be used to apprehend perpetrators, assist targets, and avoid future instances of offenses [1, 9]. Niantic enforces a 'series of disciplinary actions that gradually increase in severity and provide multiple opportunities to change their behavior' [58], however, these mainly concern in-game cheating behaviors.

While is it without doubt that further safety policies for LB/AR games need to be formulated by service providers, the difficulty lies in their implementation: what does crime prevention mean in context with both digital and physical spaces? Answers to these challenges are likely to evolve in concurrence with new legal frameworks which cater to the technological changes [13]. The security risks identified in this study inscribe themselves within the broader safety and safeguarding discussions initiated by legislation such as the UK's Online Safety Bill and the US state of California's Age-Appropriate Design Code Act. These laws place a greater responsibility on service providers to protect their users irrespective of their digital literacy.

The results from the risk assessment and quasi-experiment have generated potential implications that policy-makers and stakeholders may be interested in. However, the work presented is not without limitations. The primary of which is the lack of participants in gift-exchange (n = 4), which severely limits the generalizability of our findings. Further, the quality of data is found wanting. This is due to the limitations with advertising the quasi-experiment to obtain participants in online communities for the purposes of research. As participation was voluntary and made clear that it was for a quasi-experiment study, participants were less inclined to fully engage with Researcher 2. This behavior can be seen in Participants 2 and 5, who sent two and zero gifts respectively over the gift-exchange phase.

Despite these limitations, the risk assessment and quasi-experiment serves as a preliminary/proof-of-concept study which has shown that the risk event identified is both feasible and potentially plausible. Should stakeholders be interested, the next step would be to develop this study further. Our work focuses on how user behavior leaks technical information in a way that violates user privacy. This allows offenders to exploit leaked information and exposes the playerbase to the risk of cyber-enabled stalking. Further studies could examine similar violations of privacy in features of other LBGs and/or AR applications. By systematically identifying problematic in-game/app features, future work could then explore potential privacy-enhancing updates to the features without compromising the spirit of location-based gameplay.

## 6 CONCLUSION

The purpose of this study was to identify and analyze the potential safety risks posed by PoGO from an offender perspective and test the feasibility of its most probable risk event: cyber-enabled stalking. While this type of offence has been previously reported to have happened by users, this study is the first to provide theoretical foundations and empirical data to prove its feasibility. Our experimental results serve as a proof of concept, demonstrating the possibility for augmented reality games to be exploited maliciously by stalkers and identify victims' routines and locations. Further empirical research on PoGO is warranted to explore these new security implications especially considering Niantic's upcoming social networking platform 'Campfire' which will allow all Niantic games and players to be interconnected on one application [52].

We have focused on PoGO in this study due to Niantic's industry-leading presence. The risk events identified, specifically cyber-enabled stalking, can also be found in-game features in similar LBGs. The challenges posed by the expansion of LB/AR games reside the ambiguity and difficulty to identify offenders and prevent their actions when the virtual spaces are overlaid on physical ones. It also highlights the potential privacy-eroding features which, unlike other features, is not as straightforward to automatically detect as a geotag or similar technology. The growth of metaverse-like platforms embodies the desire to increasingly merge these environments enabling a wide range of positive game-play experiences but also opportunities for malicious exploitation since victims and offenders will intersect both on and off-line. This demands further research in responsible game design, online safety and digital literacy, in addition to risk assessments of potential exploits on LBGs and metaverse platforms.

## A  RISK SCREENING MATRIX

Table 3.  Risk screening of risk events identified through in-game features of PoGO.

| Identified Risk Event | Scenario Explanation | Severity of Consequences | Probability |
|---|---|---|---|
| *Opportunistic Events* | | | |
| **1. Theft/Robbery** | Offenders find victims at PokéStops and Gyms, perhaps during an active lure or a raid. They then commit theft or robbery depending on the use of violence/coercion or not. | Medium. This results in stolen goods (eg phone or wallet). Could be high if offenders use violence which results in injury. | High. Players focused on the game become attractive targets for theives who abuse their inattention. Highly frequented and remote areas both have high risk profiles. This has been reported [22, 76]. |
| **2. Assault/ Sexual Assault** | Offenders find victims at PokéStops and Gyms, perhaps during an active lure or a raid. They then carry out a physical assault either immediately or after a short real world conversation. | High. Assault can yield serious injury or rape/death. | Medium. Offenders and targets may coincide at PokéStops and Gyms, spontaneously engage in arguments or harassment leading to physical violence. |
| **3. Manslaughter** | Offenders find victims at PokéStops and Gyms, perhaps during an active lure or a raid. They then carry out manslaughter either immediately or after a short real world conversation. | High. Manslaughter leads to loss of life and attempted manslaughter can be seriously injuring. | Very low. Manslaughter is a relatively rare event in the population compared to other crimes considered here. |
| *Premeditated Events* | | | |
| **4. Theft/Robbery** | Offenders find victims at PokéStops and Gyms, perhaps during an active lure or a raid. They then commit theft or robbery depending on the use of violence/coercion or not. | Medium. This results in stolen goods (eg phone or wallet). Could be high if offenders use violence which results in injury. | High. Players focused on the game become attractive targets for theives who abuse their inattention. Highly frequented and remote areas both have high risk profiles. |

| **5. Assault/ Kidnapping** | Offenders find victims at PokéStops and Gyms, perhaps during an active lure or a raid. They then carry out a physical assault or kidnapping either immediately or after a short real world conversation. | High. Assault can yield serious injury or rape/death. Kidnapping results in further risks to the individual. PoGO allows offenders to lure random victims to secluded areas via lures on remote PokéStops. | Medium. Players focused on the game can become attractive targets for assault. Kidnapping is easier in more remote or more high density locations. |
|---|---|---|---|
| **6. Drug Dealing** | Offenders could deal drugs to passersby at PokéStops or Gyms; these could be further advertised on social media. Offenders may also propose battles with friends. Each Pokémon used represents a different drug for sale: eg the common Pokémon Pikachu could be cannabis and a poisonous-type like Ghastly could be heroin | High. While some drugs can be relatively benign and legal in other jurisdictions, drugs could be laced with unknown substances or overused. New users could be introduced via this social platform. | Low. PokéStops which are located in natural concentrations of people lend themselves well to drug dealing. This likely occurs independently of PokémonGO. The battle situation is improbable and likely requires another communication channel, defeating the purpose. |
| **7. Cyberstalking** | Offenders can see usernames of victims at PokéStops and Gyms or via their friends list. If victims re-use usernames across platforms, offenders can easily track them. Location can be used to confirm. | Low. This may cause stress, anxiety and harm to the victim, but there are no severe consequences identified specifically. It is recognised as a crime under harassment laws in various countries, but is not solely used to convict an offender. | Medium. With virtually unlimited interactions with other players, obtaining usernames is trivial. Recycling of usernames is required to enable this, but of medium probability. |
| **8. Cyber-enabled stalking** | The repeated exchange of gifts allows offenders to find players' most visited locations, identifying the physical areas near where they live or work, thus enabling stalking. | High. Unlike cyberstalking, cyber-enabled talking enters into the physical world. Unwanted in person attention can escalate to further crimes like assault. | High. Offenders may easily use the gift exchange feature with known or unknown players. Online tools such as maps specifically created for PokémonGO also facilitate this process [57]. |
| **9. Murder** | The repeated exchange of gifts allows offenders to find players' most visited locations, identifying the physical areas near where they live or work, which allows the offender to locate the victim and commit murder. | High. Murder leads to loss of life and attempted murder can be seriously injuring. | Low/Medium. While offenders may easily use the gift exchange feature with others and use online maps tools to facilitate this, this requires offenders to attract players to PokéStops that are isolated enough and perhaps at certain times of the day (late at night or early morning). This scenario is feasible given the location data available. |
| **10. Grooming** | The repeated exchange of gifts allows offenders to find players' most visited locations, identifying the physical areas near where they live or work, which allows the offender to locate the victim. The friends list feature enables offenders to add or invite existing contacts (potentially minors) to play PokémonGO. Grooming via gifts could lead to eventual contact and potentially escalating crimes. | High. Grooming with physical contact could lead to escalating crimes like sexual assault. | Medium. Offenders via gift exchange and trading could slowly gain the victim's trust and favour, and eventually lead to physical contact. |

| 11. Sexual Assault/ Rape | The repeated exchange of gifts allows offenders to find players' most visited locations, identifying the physical areas near where they live or work, which allows the offender to locate the victim. This could be exploited by offenders by allowing them to first add their victim through their Friends list. | High. Sexual assault/rape is an injurious, often violent crime. | Medium. While offenders may easily use the gift exchange feature with others and use online maps tools to facilitate this, this potentially requires offenders to attract players to PokéStops that are isolated enough and perhaps at certain times of the day (late at night or early morning). |
|---|---|---|---|

## B   DEFINITIONS OF IN GAME FEATURES

Table 4.   Definitions table for in-game features involved in cyber-enabled stalking.

| In-game Feature | Definition |
|---|---|
| Friend | An in app connection between two users |
| Friends List | A list of friends who are connected to a single user, max 400 friends. |
| Friendship Level | 5 can be acheived: 'Friends', 'Good Friends', 'Great Friends', 'Ultra Friends', and 'Best Friends.' Users progress through the milestones by completing interactions with a friend (eg battle, trade, gift exchange) |
| Pokémon | Characters that the game is centered around. Used in battles and trades. Found in the wild at a random location in app that corresponds to a random physical location. |
| PokéStop | A consistent location in the app that overlays onto the physical world, often at historical or notable sites. These generate items including gifts given to users at the location for use in game. |
| Gym | A consistent location in the app that overlays onto the physical world, often at historical or notable sites. Battles take place here. |
| Postcard | A visual representation of a location, inspired by Japanese train station postcards. These optionally indicate which user they were given from. No geotag, but the name is often indicative of precise location and always indicative of region. |
| Gift | An object passed between friends or from PokéStop to user. These include a postcards along with other game enhancers. Users can pick which postcard to pass along in the gift package. |
| Trainer Battle | Two users can dual their Pokémon in person (if a low or no friendship level) or remotely (if Ultra or Best Friends). Users earn points for winning the battle and in-game objects for participating. |
| Gift Exchange | A user can remotely share a gift with a friend, max 1 per day per friend. The postcard comes out of the user's collection while the rest of the gift is generated randomly and not taken from users' items. |
| Trade | Two friends can meet up in a proximal physical location at the same time to swap Pokémon from their collections. |
| Raids | A group Pokémon hunting experience that takes place at point of time at a Gym. Users need not be friends as long as they are proximal in location. |
| Lures | A game enhancer that attracts more Pokémon to a PokéStop. Everybody can see which user played the lure when proximal to the PokéStop. |

## C  USER SURVEY

Table 5.   Questionnaire given to the participants by Researcher 1.

| Question number | Question |
|---|---|
| Q1 | Please state your Pokémon Go ID: |
| Q2 | Provide your Pokémon Go friend code: |
| Q3 | Age: |
| Q4 | What are the names of the three closest Pokestops to your home? |
| Q5 | What are the names of the three closest Pokestops to your workplace? |
| Q6 | Please list the street names of up to three leisure locations you regularly attend. |
| Q7 | On average, how many days a week do you play Pokemon Go? |
| Q8 | On average, how long do you play PokemonGo per session in hours? (use decimals for less than an hour) |
| Q9 | How many friends do you have in the game? |
| Q10 | Do you know EVERYONE on your friend's list personally outside of the application? |